\documentclass[10pt,aps,prl,raggedbottom,longbibliography,reprint,citeautoscript,letterpaper,superscriptaddress]{revtex4-2}
\usepackage[usenames,dvipsnames]{color}
\usepackage{graphicx,microtype,lmodern}
\graphicspath{{./}{Figures/}}
\usepackage{multirow,amsmath} 
\usepackage[bookmarks=false,colorlinks]{hyperref}
\hypersetup{linkcolor=magenta,citecolor=MidnightBlue,filecolor=Plum,urlcolor=MidnightBlue}

\usepackage{easyReview}
\usepackage{amssymb}



\newcommand{\supf}{\textcolor{DarkOrchid}{Fig.}\,S}
\newcommand{\supt}{\textcolor{DarkOrchid}{Table}\,S}

\newcommand{\rev}[1]{\textcolor{RoyalBlue}{#1}}

\begin{document}

\title{Dynamical Nonrelativistic Spin Splitting via THz Nonlinear Phononics}

\author{Linding Yuan}
\email{linding.yuan@northwestern.edu}
\affiliation{Department of Materials Science and Engineering, Northwestern University, Evanston, Illinois 60208, USA}

\author{ChanJu You}
%\email{cy454@cornell.edu}
\affiliation{Department of Physics, Cornell University, Ithaca, NY, USA}

\author{Ankit Disa}
\email{asd47@cornell.edu}
\affiliation{School of Applied \& Engineering Physics, Cornell University, Ithaca, NY, USA}

\author{James M.\ Rondinelli}%
\email{jrondinelli@northwestern.edu}
\affiliation{Department of Materials Science and Engineering, Northwestern University, Evanston, Illinois 60208, USA}

%\date{\today}

\begin{abstract}
Nonrelativistic spin splitting (NRSS) in collinear antiferromagnets, including altermagnets and compensated ferrimagnets, offers a route to high-frequency spintronics immune to stray fields, but its dynamic control has remained elusive. 
We demonstrate, using density functional theory (DFT) and nonlinear phononics, that THz laser pulses can achieve ultrafast, reversible control of NRSS on picosecond timescales in antiferromagnets.
We derive two mode-selection criteria, accounting for phonon and magnetic wavevector compatibility and inversion parity, to identify which phonon modes can activate NRSS. 
Our calculations illustrate the creation of NRSS in spin-degenerate NiO and the enhancement of existing splitting in LaFeO$_3$.
While practical implementation depends on excitation conditions, lattice dynamics, and mode coupling, these results establish design principles for identifying materials with stronger and more accessible responses. Limitations imposed by driving fields, phonon damping, and anharmonic detuning motivate further materials screening based on accessible excitation strengths, lattice stability, resonance frequency matching, and favorable anharmonic coupling.
\end{abstract}

\maketitle

\paragraph*{Introduction.} Crystal structure serves as the picoscale blueprint that governs the electronic properties of a compound by encoding atomic composition, bonding and symmetry, which together determine energy--momentum dispersions (band structure) \cite{ismail-beigi_picoscale_2017,zunger2018inverse,10.1063/1.4928289}. 
In collinear antiferromagnets with suitable atomic arrangements and crystalline symmetries, this interplay can yield large symmetric (even-parity) nonrelativistic spin splitting (NRSS) driven purely by exchange, without relying on relativistic spin--orbit coupling (SOC) \cite{noda2016momentum,Naka2019,doi:10.7566/JPSJ.88.123702,PhysRevB.99.184432,PhysRevB.102.014422,PhysRevB.101.220403,PhysRevB.102.144441,doi:10.1126/sciadv.aaz8809,PhysRevMaterials.5.014409,ma2021multifunctional,PhysRevX.12.031042}. 
Such materials, dubbed as altermagnets, are promising for spintronics, because they combine large spin splitting with fast dynamics and no stray fields \cite{fender2025altermagnetism,song2025altermagnets,jungwirth_altermagnetic_2026}. 

To date, NRSS control has largely relied on static structural modifications, such as strain or chemical substitution \cite{zhou_manipulation_2025}, which are inherently limited by thermodynamic and elasticity constraints while also lacking temporal control.
An emerging alternative is dynamic structural control, in which lattice degrees of freedom are driven far from equilibrium through mode-selective optical excitation \cite{forst_nonlinear_2011,PhysRevLett.118.054101,disa_engineering_2021,PhysRevB.109.024110}.
In this framework, mid-infrared or THz pulses coherently excite infrared-active (IR) phonons, which anharmonically couple to other modes of the system and induce transient atomic displacements along that coordinate.
This nonlinear phononics (NLP) mechanism enables coherent coupling between photons and phonons, ultrafast manipulation of crystal symmetry and functionality, often accessing states unattainable in equilibrium.
Notable demonstrations include light-induced enhancement of superconductivity \cite{fausti2011light,mankowsky_nonlinear_2014}, optical control of magnetism \cite{disa_polarizing_2020,stupakiewicz_ultrafast_2021,doi:10.1126/science.adi9601,disa_photo-induced_2023,davies_phononic_2024}, light-induced ferroelectricity and its switching\cite{PhysRevB.97.085145,PhysRevLett.118.197601}, and photo-induced chirality and ferroaxiality \cite{doi:10.1126/science.adr4713,doi:10.1126/science.adz5230}.

While optical control over exchange interactions and magnetization have been previously explored~\cite{khalsa_ultrafast_2018,disa_polarizing_2020,stupakiewicz_ultrafast_2021,doi:10.1126/science.adi9601,disa_photo-induced_2023,davies_phononic_2024,PhysRevMaterials.2.064401,doi:10.1126/sciadv.ado0722}, whether nonlinear phononics can be systematically harnessed to control NRSS via lattice modulations remains an open question.
Early first-principles studies on perovskite titanates and manganites \cite{gu_ultrafast_2016}, and more recently MnPS$_3$ \cite{PhysRevB.108.054308} and MnF$_2$ \cite{c3kh-5s5t}, showed that nonequilibrium lattice excitations can induce transient spin splitting in antiferromagnets, but the governing principles were articulated only at a heuristic level and no general design rules have been established.

\begin{table*}[t]
  \caption{Symmetry breaking modes that activates NRSS in representative collinear antiferromagnets. Columns from left to right list the magnetic compound, parent space group (SG), magnetic irrep, time-reversal preserving symmetry breaking mode, NRSS active mode criteria 1 and 2, and whether NRSS is activated. A-AF, C-AF, G-AF denote A-type, C-type, and G-type AFM configurations, respectively.}
  \label{tab:nrss_conditions}
  \squeezetable
  \setlength{\tabcolsep}{4pt}
  \begin{ruledtabular}
  \begin{tabular}{llllccc}
    Compound & Parent SG & Mag. irrep & Symmetry-breaking mode &
    \shortstack{Criteria 1\\compatible $\mathbf{k}$} &
    \shortstack{Criteria 2\\matching parity} &
    \shortstack{NRSS\\activated} \\
    \hline
    MnPS$_3$~\cite{rajpurohit2026symmetry} &
    $C2/m$ & $\mathrm{m}\Gamma_{2}^{-}$ &
    $\Gamma_{1}^{-}$, $\Gamma_{2}^{-}$ (resonant IR) &
    $\checkmark$ & $\checkmark$ & Yes \\

    Ca$_2$MnO$_4$~\cite{zhang_tunable_2026} &
    $I4_{1}/acd$ & $\mathrm{m}\Gamma_{4}^{-}$ &
    $\Gamma_{2}^{-}$ ($E$\mbox{-}field~$\parallel z$) &
    $\checkmark$ & $\checkmark$ & Yes \\

    LaMnO$_3$ (A-AF)~\cite{https://doi.org/10.1002/adma.202211966} &
    $Pnma$ & $\mathrm{m}\Gamma_{4}^{+}$ &
    $\Gamma_{1}^{+}$ (no distortion) &
    $\checkmark$ & $\checkmark$ & Yes \\

    LaMnO$_3$ (C-AF)~\cite{https://doi.org/10.1002/adma.202211966}  &
    $Pnma$ & $\mathrm{mZ}_{1}^{-}$ &
    $\Gamma_{1}^{+}$ (no distortion) &
    $\times$ & $\times$ & No \\

    KCuF$_3$ (A-AF)~\cite{han2026strain} &
    $I4/mcm$ & $\mathrm{mM}_{5}^{+}$ &
    $\Gamma_{1}^{+}$ (no distortion) &
    $\times$ & $\checkmark$ & No \\

    KCuF$_3$ (G-AF)~\cite{han2026strain} &
    $I4/mcm$ & $\mathrm{m}\Gamma_{5}^{+}$ &
    $\Gamma_{1}^{+}$ (no distortion) &
    $\checkmark$ & $\checkmark$ & Yes \\
  \end{tabular}
  \end{ruledtabular}
\end{table*}

In this Letter, we combine first-principles calculations, group-theoretical analysis, and effective-mode modeling to 
establish a symmetry-based framework for ultrafast control of NRSS via light-driven lattice excitations (\supf 1). %(\autoref{fig:thz-symmetry-control}).
We formulate practical selection rules showing that a driven phonon activates  NRSS when its wavevector is compatible with the magnetic propagation vector and its inversion parity matches that of the magnetic order, transforming a heuristic search into a predictive process for spin control.
We apply these rules to NiO and LaFeO$_3$, identified the distortion modes and verified with Density functional theory (DFT) that creates NRSS from a spin degenerate state or modify pre-existing altermagnetic splitting, respectively.
We then discuss experimental signatures, feasibility, and detection of the dynamical effects.
Our analysis of NiO and LaFeO$_3$ uncover limitations  arising from the required field strength, picoscale displacements, phonon damping, and anharmonic detuning; We therefore view NiO and LaFeO$_3$ as proofs of principle rather than established experimental platforms.
Quadratic-linear coupling and biquadratic couplings that soften a Raman mode can offer a picosecond metastable NRSS state but demand careful materials selection.
These findings motivate further theoretical materials screening: identify symmetry modes that activates NRSS, evaluate their coupling to optically accessible phonons (if they are not directly accessible), and assess whether the resulting dynamics can generate appreciable NRSS with realistic damping and lattice-stability limits.

\paragraph*{Symmetry rule.} In the absence of SOC, spin degeneracy of the bands across the Brillouin zone in collinear antiferromagnets is protected by two combined symmetries: time-reversal combined with spatial inversion ($\Theta I$), and spin rotation followed by a lattice translation ($U\tau$) \cite{PhysRevB.102.014422,PhysRevMaterials.5.014409}.
Achieving NRSS thus requires the simultaneous breaking of both symmetries.
For spin-degenerate antiferromagnets, this is  accomplished by driving a symmetry breaking mode $Q$ that lifts these protections \cite{1r6k-s46h}.

\begin{figure*}[t]
  \centering
  \includegraphics[width=\linewidth]{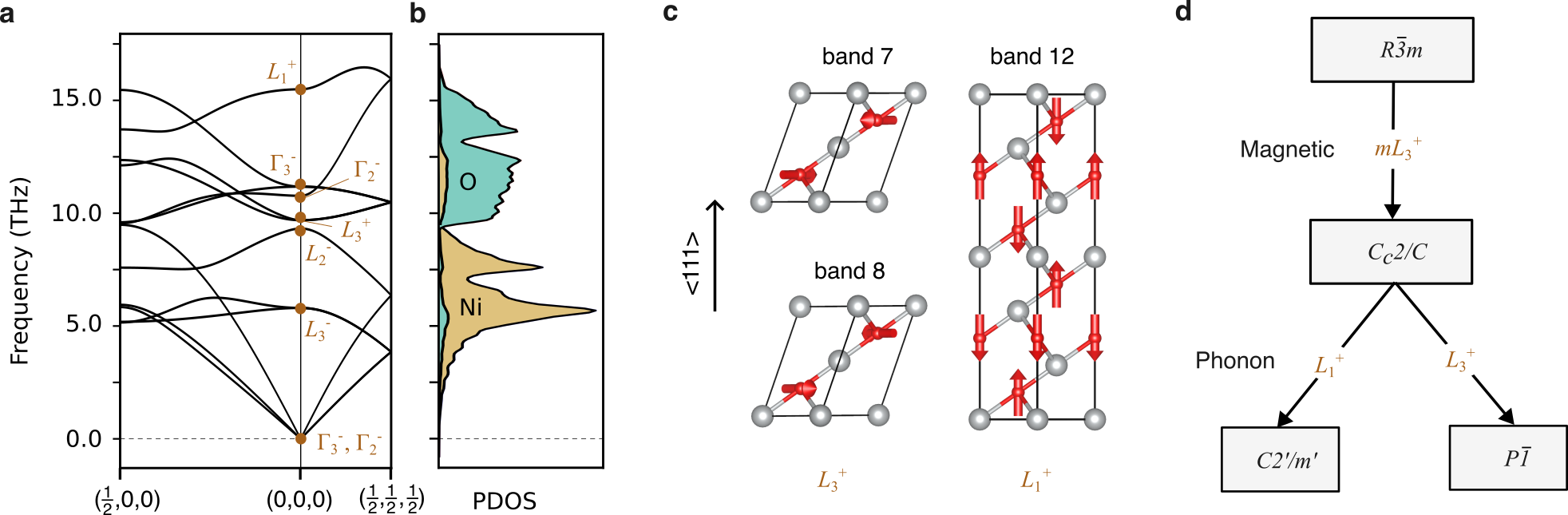}\vspace{-6pt}
  \caption{Calculated NiO (a) phonon dispersions, (b) atom-projected phonon density of states, (c) NRSS-enabling phonon modes $L_3^+$ and $L_1^+$, and (d) the associated symmetry mapping. In (c), Ni and O atoms are represented by gray and red spheres, respectively; atomic displacement directions are indicated by red arrows. %LO/TO splitting (
  (Nonanalytic corrections for $q \to 0$ are ignored.)}
  \label{fig:phonon-nio}
\end{figure*}

To translate this requirement into a practical mode selection criterion, we recast the symmetry conditions in terms of magnetic irreducible representations (irreps) of the parent paramagnetic space group \cite{bertaut1968representation,opechowski1965magnetic}.
This approach allows direct connection to phonon irreps, e.g., $Q_\Gamma=\Gamma_4^+$, which label the driven normal modes as irreps of the crystallographic space group and  cannot distinguish reductions in $\Theta I$ and $U\tau$ from $Q_\Gamma$. %\ad{What does this sentence mean? Does it mean that breaking these symmetries does not change the irrep?}
The resulting selection rules (see the Supplementary Materials, SM \cite{supp}) are:
($i$) the phonon wavevector $\mathbf k_Q$ must be compatible with the magnetic propagation vector $\mathbf k_M$, such that the induced distortion removes the translation component underlying $U\tau$; equivalently, it removes the ``anti-translation'' by converting even-denominator components of $k_M$ into odd denominators in the distorted supercell \footnote{Propagation vectors are often reported in reciprocal coordinates of the conventional crystallographic cell, which may be non-primitive. Integer components therefore do not necessarily indicate a zone-center mode; for example, ($\mathbf k_M=(1,0,0)$) is a zone-boundary vector for a body-centered cubic lattice. Wavevector compatibility must be checked against all lattice translations, including centering translations. The even-/odd-denominator test should be applied in a primitive reciprocal basis of the distorted structure.}. This condition is satisfied when the phonon and magnetic order have the same propagation vector $\mathbf k_Q=\mathbf k_M$. But equality is not necessary; for example, in a chain with lattice spacing ``a'' and $\mathbf k_M=(1/6,0,0)$, a distortion with $\mathbf k_Q=(1/2,0,0)$ also satisfies this requirement. Its structural period of ``2a'' eliminates the odd-multiple-of-a translation that exchanges opposite spins.
($ii$) When inversion symmetry is relevant, the phonon must share the same inversion parity as the magnetic order (both being ``+'' or ``-''). It is worth noting that a ``-'' phonon mode does not necessarily remove all the structural inversion and break $\Theta I$ symmetry. For example, a ``-'' zone boundary mode removes the inversion center at the origin but preserves inversion about a different location; when the parity of the magnetic order is opposite (i.e., ``+''), $\Theta I$ symmetry is preserved. Matching parity ensures that the combined antiunitary symmetry $\Theta I$ is removed.

To illustrate the selection rules, we applied them to the analysis of a few example materials. The results are summarized in \autoref{tab:nrss_conditions} and are consistent with the aforementioned criteria. For example, an IR-active zone-center mode with odd inversion parity ($\Gamma_X^-$) in MnPS$_3$ (whose magnetic order transforms as $m\Gamma_2^-$) satisfies the two criteria and induces NRSS~\cite{rajpurohit2026symmetry}.
The applicability of our theory extends to other time reversal preserving symmetry breaking (SB) modes, such as an electric field. It can also be applied to identify NRSS AFMs by considering the SB mode (no distortion) as a $\Gamma_1^+$ or $\Gamma_1$ mode; thus, an antiferromagnet with a magnetic irrep of $m\Gamma_X^+$ that has a compatible-$k$ and matching parity to $\Gamma_1^+$ will exhibit NRSS. In fact, most  encountered altermagnets have $m\Gamma_X^+$  magnetic irreps, e.g., MnF$_2$ ($m\Gamma_2^+$), MnTe ($m\Gamma_5^+$), alpha-Fe$_2$O$_3$ ($m\Gamma_3^+$). These examples illustrate the general applicability of our mode-selection rules.

\paragraph*{Application to NiO.}
To demonstrate the NLP strategy, we consider the prototypical spin-degenerate rock-salt antiferromagnet NiO \cite{shull1951neutron}. 
In equilibrium, NiO comprises ferromagnetic $(111)$ planes stacked antiferromagnetically, lowering symmetry from $Fm\bar{3}m$ to the magnetic space group $C_c2/c$ with active magnetic irrep $mL_3^+$ \cite{baruchel1981antiferromagnetic}.
The associated propagation vector $L=(1/2,1/2,1/2)$ contains even fractional denominators, preserving the  $U\tau$ symmetry and enforcing complete spin degeneracy across the BZ.
NRSS can be induced by a phonon that  breaks both $\Theta I$ and $U\tau$.
A distortion at the $L$-point of the parent paramagnetic structure doubles the unit cell along $[111]$, folding the active magnetic irrep ($mL_3^+$) %from $L$ 
to $\Gamma$ and eliminating the translation symmetry underlying $U\tau$.
Thus, the $L$-point distortions remove the even denominators from the magnetic propagation vector, and its unit-cell doubling is not accessible via static strains. %\ad{I find the discussion about the denominator of the propagation vector confusing. Would it be easier to just talk about the fact that the magnetic space group has a doubled unit cell, and then pumping a mode with the same wavevector allows for coupling to the magnetic order?}
Simultaneously, breaking $\Theta I$, requires that the phonon carry the same even inversion parity (``$+$'') as the active magnetic irrep.

\paragraph*{First-principles validation.}
We evaluate the predictions using DFT calculations
\cite{supp}.
%(see SM for methods \cite{supp}).
%
The calculated phonon spectrum of NiO 
(\autoref{fig:phonon-nio}a), yields 12 zone-center modes for the four-atom magnetic unit cell. 
\autoref{tab:nio-lattice-dynamics} lists the optical modes, including their frequencies, irreps, and associated magnetic symmetry constraints on NRSS. 
Although labeled at the magnetic-zone center, several modes originate from $L$-point irreps of the parent unit cell and  fold to $\Gamma$ in the magnetic unit cell.
Among the nine optical modes, only two Raman-active modes $L_3^+ (7,8)$ and $L_1^+ (12)$ satisfy the parity-matching condition. Both are dominated by opposing oxygen displacements (\autoref{fig:phonon-nio}b,c), while the Ni atoms remain nearly fixed.
(Hereafter, we drop the band index for the $L$ modes for compactness, unless there is ambiguity.)

\begin{figure*}
  \centering
  \includegraphics[width=\linewidth]{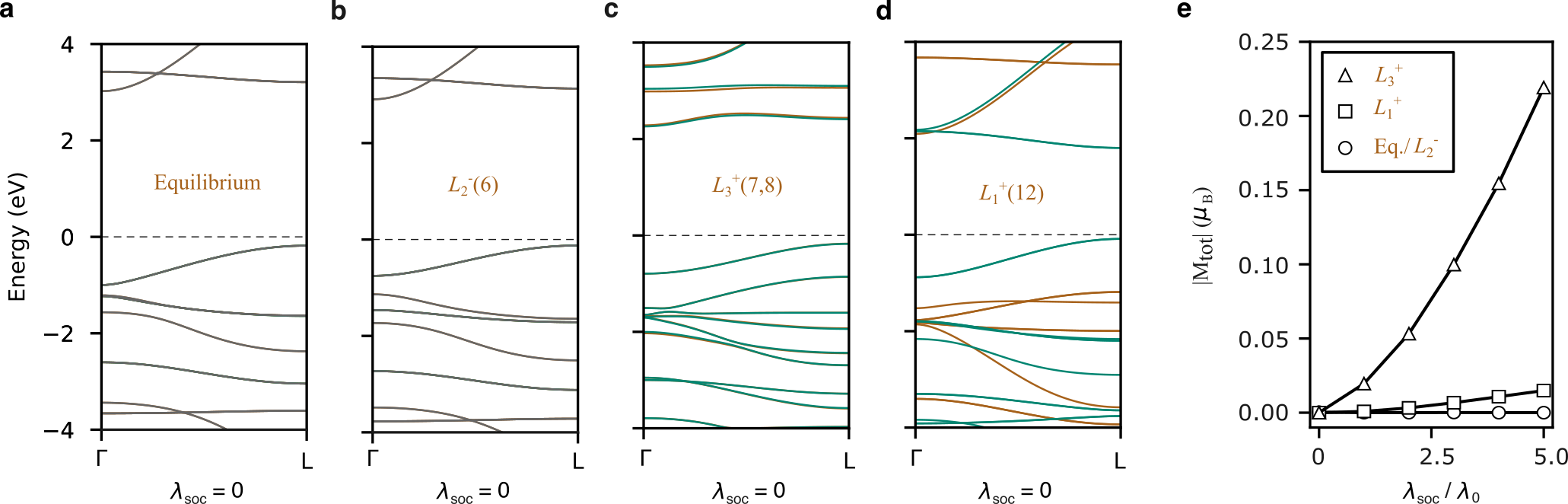}%\vspace{-4pt}
  \caption{NiO electronic band dispersions for (a) undistorted and quasi-static structures modulated with (b) $L_2^- (6)$ phonon, (c) $L_3^+ (7,8)$ phonon, and (d) $L_1^+ (12)$ phonon. Spin-up and spin-down bands are indicated in jade and brown, respectively.
  (e) SOC strength dependence of the net magnetization for the equilibrium (Eq.) and phonon-mode-modulated NiO structures.  Phonon mode amplitude $|Q|$ is set to 1 \AA$ \sqrt{\mathrm{amu}}$ in all cases.}
  \label{fig:dft-nio}
\end{figure*}

The spin-polarized band structure of equilibrium NiO (without SOC) is doubly degenerate (\autoref{fig:dft-nio}a, gray lines), consistent with our symmetry analysis.
A distortion along the $L_2^-$ mode (amplitude $|Q|=1.0\,\text{\AA}\sqrt{\mathrm{amu}}$) preserves this degeneracy (\autoref{fig:dft-nio}b), as its odd parity ($-$) is incompatible with the even parity ($+$) of the active magnetic irrep.
In contrast, distortions along $L_3^+$ or $L_1^+$ (same amplitude) lift the spin degeneracy throughout the BZ (\autoref{fig:dft-nio}c,d).
The splitting persists at $\Gamma$, placing these nonequilibrium states in the compensated-ferrimagnet class rather than the altermagnet regime, which requires degeneracy at $\Gamma$.

The emergence of NRSS at $\Gamma$ can be traced to bond disproportionation between neighboring edge-sharing NiO$_6$ octahedra along $\langle111\rangle$.
Both $L_1^+$ and $L_3^+$ split the Ni Wyckoff orbit (4a) into two inequivalent sublattices (3a,3b) with opposite spin orientations, removing all crystal symmetries (including mirror reflections) and their combinations with $\Theta$ that would relate them.
This symmetry breaking generates a non-interconvertible spin-structure motif pair \cite{yuan2023degeneracy}, providing the structural origin of NRSS at $\Gamma$.
Although the lowered magnetic symmetry of the distorted structures allows a ferromagnetic component in principle, our nonrelativistic calculations (SOC off) yield strictly compensated moments. The two inequivalent Ni sublattices carry equal and opposite integer spin moments, so the total magnetization remains zero while the band degeneracy at $\Gamma$ is lifted.
When SOC is included, however, both $L_1^+$ and $L_3^+$ modulated structures acquire a small but finite net moment (\autoref{fig:dft-nio}e); for NiO this SOC-induced moment is weak compared with the exchange-driven NRSS.

\begin{table}[b]
  \caption{Zone-center optical phonon modes in NiO from DFPT calculations (differ slightly from the frozen phonon mode calculations) with magnetic irrep $mL_3^+$ and their impact on NRSS under a driven-IR mode. Mode effective charges are in units of $e\textrm{\AA}\sqrt{\textrm{amu}}$.}
  \label{tab:nio-lattice-dynamics}
  \begin{ruledtabular}
  \begin{tabular}{lllllll}
    Band & $\omega$ (THz) & SG & MSG & IR $Q_\Gamma$ & $|Z^*|$ & NRSS? \\
    % &(THz)&&&& ($e\mathrm{\AA}$)& \\
    \hline
    %$1,2$ & 0 & $R-3m$ & $C_c2/c$ & $\Gamma_2^{-}$ & 0 & No\\
    %$3$ & 0 & $R-3m$ & $C_c2/c$ & $\Gamma_3^{-}$ & 0 & No\\
    $4,5$ & 5.77 & $P\bar{1}$ & $P\bar{1}^{\prime}$ & $L_3^{-}$ & 0 & No \\
    $6$ & 9.35 & $R\bar{3}m$ & $C2/m^{\prime}$ & $L_2^{-}$ & 0 & No \\
    $7,8$ & 9.65 & $P\bar{1}$ & $P\bar{1}$ & $L_3^{+}$ & 0 & Yes \\
    $9$ & 10.74 & $R3m$ & $C_cc$ & $\Gamma_2^{-}$ & 0.88 & No \\
    $10,11$ & 11.08 & $P1$ & $P_S1$ & $\Gamma_3^{-}$ & 0.86 & No \\
    $12$ & 15.57 & $R\bar{3}m$ & $C2^{\prime}/m^{\prime}$ & $L_1^{+}$ & 0 & Yes
  \end{tabular}
  \end{ruledtabular}
\end{table}

\paragraph*{Nonlinear phononics.} Within the NLP framework, lattice dynamics are controlled by selectively driving normal-mode coordinates with resonant light pulses \cite{subedi_theory_2014}.
A THz field excites an IR-active phonon, which subsequently transfers energy to Raman-active modes through anharmonic interactions, thereby generating symmetry-lowering atomic-scale distortions.
We note that in strongly polar/ionic crystals the driven IR phonon can hybridize with the THz electromagnetic field, forming propagating phonon--polaritons that can enhance both optical and lattice nonlinearities and transiently renormalize electronic structure \cite{lu_nonlinear_2024,lu_stimulated_2022}.
Our NLP model assumes a spatially uniform driven coordinate and therefore neglects polariton propagation and field inhomogeneity; including such effects could quantitatively modify  the predicted NRSS response.
As the NRSS effect induced by the $L_3^+$ mode is significantly weaker than that of $L_1^+$, we focus on selective excitation of the $L_1^+$ Raman mode. 
To identify the relevant IR modes that strongly couple to the Raman modes, we compute Born effective charges (BECs) 
%using density functional perturbation theory 
and evaluate the corresponding mode effective charges $Z^*$.
Among the nine optical modes (\autoref{tab:nio-lattice-dynamics}), only two IR-active modes, $\Gamma_2^-(9)$ and $\Gamma_3^-(10,11)$, have $Z^*\neq0$ and are thus directly excitable by THz radiation.

\begin{figure*}
  \centering
  \includegraphics[width=0.93\linewidth]{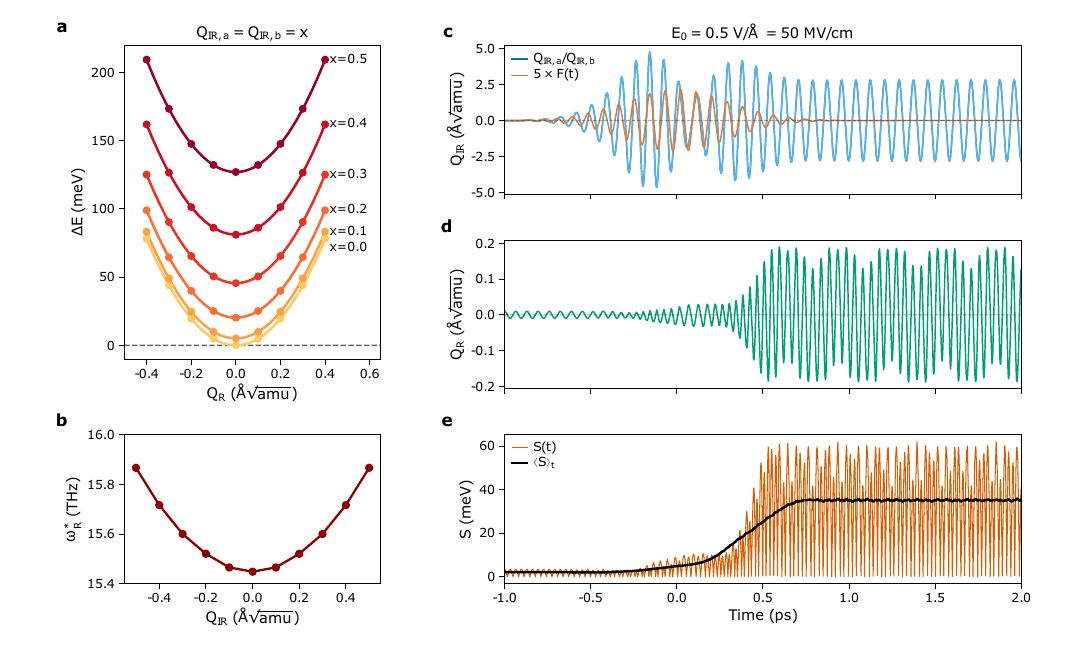}\vspace{-11pt}
  \caption{Energy surface and phonon dynamics of NiO. (a) Potential energy profiles for the nonlinear coupling between $\Gamma_3^-(10,11)$ and $L_1^+ (12)$ modes as a function of $Q_\mathrm{R}$ at fixed IR amplitudes $Q_{\mathrm{IR,a}} = Q_{\mathrm{IR,b}} = x$. (b) Modulated Raman-mode frequency as a function of IR amplitude $x$. Time evolution of the (c) laser pulse and the two components of the $\Gamma_{3}^- (10,11)$ IR mode; (d) $L_1^+ (12)$ Raman mode; and (e) transient $S$ and time-averaged (over 0.5\,ps period) mean $\langle S \rangle_t$ spin splitting.}
  \label{fig:eom-nio}
\end{figure*}

Two anharmonic cubic ($Q^2_\mathrm{IR} Q_\mathrm{R}$) and quartic ($Q^2_\mathrm{IR} Q^2_\mathrm{R}$) coupling channels are relevant; however, cubic coupling between the $\Gamma$ IR modes and the $L$ Raman mode is strictly forbidden in NiO.
The anti-translation symmetry (comprising  time-reversal combined with translation) reverses the sign of the cubic coupling term, forcing its coefficient to vanish. Thus, the minimal lattice potential energy is $V(Q_\mathrm{IR},Q_\mathrm{R}) = 1/2\, \Omega_\mathrm{IR} Q_\mathrm{IR}^2 + 1/2 \,\Omega_\mathrm{R} Q_\mathrm{R}^2 + \rev{1/2}\,gQ_\mathrm{IR}^2\,Q_\mathrm{R}^2$. This is confirmed by DFT energy surfaces %$E(Q_\mathrm{R}, Q_\mathrm{IR})$ 
that show a negligible cubic coefficient %($\sim 10^{-18}$eV\,\AA$^{-3}$\,amu$^{-3/2}$) 
and a finite quartic coupling strength (Table \ref{tab:fit_coeffs}).
Consequently, the IR-to-Raman energy transfer is governed by quartic interactions in NiO.
%
%Among the IR modes, $\Gamma_2^-(9)$ couples only weakly to $L_1^+$ ($g=0.006$\,eV/(\AA$\sqrt{\mathrm{amu}})^4$), while $\Gamma^-_3(10,11)$ moderately couples ($g=0.106$\,eV/(\AA$\sqrt{\mathrm{amu}})^4$) and therefore serves as the primary channel for the driven dynamics we examine next. \ad{I would move the numbers to the next section after introducing the potential energy as I have in my comment.}

\paragraph*{Driven phonon dynamics.} We solve the coupled equations of motion for the IR-Raman system to demonstrate that resonant excitation of  the $\Gamma_3^-$ mode induces coherent oscillations of the  $L_1^+$ Raman coordinate via quartic anharmonic coupling, thereby generating  transient picosecond symmetry breaking. 
The driven IR coordinates satisfy 
\begin{align}
  \ddot Q_{\mathrm{IR,a}} 
  %+ 2\gamma_{\mathrm{IR}}\dot Q_{\mathrm{IR,a}} %damping term
  + g\, Q_\mathrm{R}^2 Q_{\mathrm{IR,a}} 
  + \Omega_{\mathrm{IR}} Q_{\mathrm{IR,a}} 
  &= Z^* F(t),\\
    \ddot Q_{\mathrm{IR,b}} 
    %+ 2\gamma_{\mathrm{IR}}\dot Q_{\mathrm{IR,b}} 
    + g\,  Q_\mathrm{R}^2 Q_{\mathrm{IR,b}} 
    + \Omega_{\mathrm{IR}} Q_{\mathrm{IR,b}}
  &= Z^* F(t),
\end{align}
while the Raman coordinate evolves as 
\begin{equation}
  \ddot Q_{\mathrm{R}} + %2\gamma_{\mathrm{R}}\dot Q_{\mathrm{R}} + 
  \Bigl(\Omega_{\mathrm{R}} + g\, Q_{\mathrm{IR,a}}^2 + g Q_{\mathrm{IR,b}}^2 \Bigr) Q_{\mathrm{R}}
  = 0,
\end{equation}
Here $Q_{\mathrm{IR,a}}$ and $Q_{\mathrm{IR,b}}$ denote the two components of the two-dimensional $\Gamma_3^-$ mode, $Q_{\mathrm{R}}$ is the $L_1^+$ Raman coordinate, $\Omega_{\mathrm{IR}}$ and $\Omega_{\mathrm{R}}$ are the corresponding mode stiffness (frequencies square), and $g$ is the leading quartic coupling coefficient. 
Among the IR modes, $\Gamma_2^-(9)$ couples only weakly to $L_1^+$ ($g=0.006$\,eV/(\AA$\sqrt{\mathrm{amu}})^4$), while $\Gamma^-_3(10,11)$ moderately couples ($g=0.106$\,eV/(\AA$\sqrt{\mathrm{amu}})^4$) and therefore serves as the primary channel for the driven dynamics.
% 
%For simplicity, damping and anharmonic detuning of the IR and Raman modes are neglected (see \supf1 for DFT results with damping, \supf2 for DFT results with quartic anharmonic detuning.).
%

\autoref{fig:eom-nio} shows the NLP dynamics under mid-infrared excitation. The anharmonic energy surface exhibits a systematic hardening of the $L_1^+$ potential with increasing IR amplitude ($x$ values, \autoref{fig:eom-nio}a), as the biquadratic term $g Q_\mathrm{IR}^2 Q_\mathrm{R}^2$ modifies the effective mode stiffness of the $Q_\mathrm{R}$ mode ($\Omega_\mathrm{R} \rightarrow \Omega_\mathrm{R} + gQ_\mathrm{IR}^2$).
%\ad{this should be moved to the end of the paragraph}
%
%Under a Gaussian-enveloped driving field $F_t=Z^*\,E_0\sin(\omega t)\exp({-t^2/2\sigma^2)}$ with $E_0=50$\,MV/cm and $\sigma=0.3$\,ps, the effective Raman frequency $\omega^*_\mathrm{R}$  transiently crosses $2\omega_\mathrm{IR}=22.25$\,THz, such that the two orthogonal components of the $\Gamma^-_3$ mode oscillate coherently (\autoref{fig:eom-nio}c), parametrically driving the Raman coordinate.
%\ad{Suggested replacement text: 
We consider a Gaussian-enveloped driving field $F_t=Z^*\,E_0\sin(\omega t)\exp({-t^2/2\sigma^2)}$ with $E_0=50$\,MV/cm and $\sigma=0.3$\,ps, and $\omega_{pump}$ = 11 THz, resonant with the $\Gamma^-_3$ mode, polarized to drive the two orthogonal components with equal amplitude. The drive induces coherent oscillations of the IR mode coordinates (\autoref{fig:eom-nio}c), which parametrically modulates the effective Raman frequency $\omega_R$.
%}
%
The resulting $L_1^+$ dynamics display coherent oscillations at $\sim$2$\omega_\mathrm{IR}=22.25$ THz (\autoref{fig:eom-nio}d), indicating efficient nonlinear up-conversion, a hallmark of the quartic excitation and evidence for selective Raman mode activation.
The corresponding Fourier spectrum shows a distinct peak above the bare $L_1^+$ eigenfrequency (\supf2 in the SM \cite{supp}).

\autoref{fig:eom-nio}e tracks the resulting spin splitting $S$, defined as the $k$-point-weighted mean absolute difference between spin-up and spin-down eigenvalues over all occupied bands, $S = N_{\mathrm{occ}}^{-1} \sum_{n \in \mathrm{occ}} \sum_k w_k |E_{\uparrow}(n,k) - E_{\downarrow}(n,k)|$. 
While the metric $S$ may not directly capture features near the Fermi level $E_F$, it provides a robust measure of the average spin splitting in an insulator and qualitatively correlates with magneto-optical responses (see below).
We find that the spin splitting is largely insensitive to the instantaneous IR amplitude and instead follows the Raman dynamics, oscillating at twice the Raman frequency as expected for biqudratic $Q_\mathrm{R}^2Q_\mathrm{IR}^2$ coupling.
After an initial delay during which the Raman amplitude accumulates, $S$ rises sharply and reaches 60\,meV at $\sim$0.5\,ps coincident with maximal Raman displacement.
It subsequently reaches a time-averaged value $\langle S \rangle_t \approx 40$\,meV, demonstrating a persistent spin splitting driven by NLP.

\begin{figure}
  \centering
  \includegraphics[width=\linewidth]{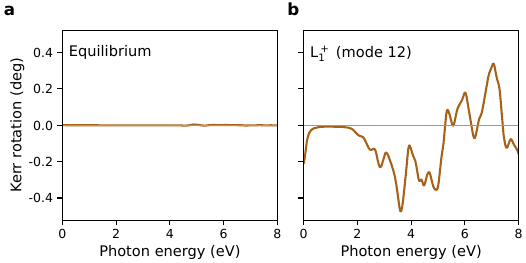}
  \caption{DFT-calculated MOKE response for light incident on the ($11\bar{2}$) surface of (a) undistorted, equilibrium NiO and (b) $L_1^+ (12)$-mode-modulated NiO at $|Q|=1.0$\,\AA$\sqrt{\mathrm{amu}}$.}
  \label{fig:moke-nio}
\end{figure}

\paragraph*{Experimental signatures and feasibility.} 
Our first-principles calculations predict that the $L_1^+$-modulated NiO exhibits pronounced spin splitting along $\Gamma$--$L$, driven by quartic anharmonic coupling between the 11.08\,THz $\Gamma_3^-$ IR mode and the 15.57\,THz $L_1^+$ Raman mode.
The associated lifting of spin degeneracy at $\Gamma$ is expected to induce a transient net ferromagnetic moment.
This moment can be probed % is in principle detectable 
via the time-resolved magneto-optical Kerr effect (MOKE) in a polar geometry, using collinear pump (11\,THz) and probe pulses incident on the ($11\bar{2}$) surface.
\autoref{fig:moke-nio} shows the DFT-calculated MOKE response.
In equilibrium, the signal vanishes due to global spin degeneracy throughout the BZ, whereas resonant excitation of the $L_1^+$ Raman mode ($|Q|=1$\,\AA$\sqrt{\mathrm{amu}}$) produces a sizeable transient Kerr rotation of up to $\sim$0.5$^\circ$, providing a clear signature of the dynamical NRSS (see also \supf3).

Under a pump field of 50\,MV/cm, the driven $L_1^+$ Raman maximum amplitude reaches $\sim$0.2\AA$\sqrt{\mathrm{amu}}$, yielding a Kerr rotation reduced by the quartic scaling to $\sim$0.02$^\circ$ oscillates at $\omega_\mathrm{R}$. 
Although the field strength exceeds typical mid-IR conditions and the MOKE signal is correspondingly small due to the quadratic scaling with the peak electric field of the pump, detection remains feasible with high-sensitivity MOKE setups.
Resolving the $\sim$22\,THz Raman oscillations further requires 
sub-100-fs temporal resolution, which is within reach of ultrafast THz spectroscopy techniques.

In practice, three effects limit feasibility. 
First, satisfying the 2:1 parametric-resonance condition (pump-induced hardening to $2\omega_{\mathrm{IR}}\approx 22$ THz) requires a 50\,MV/cm field (\supf2), driving $Q_{\mathrm{IR}}$ to $\sim$0.75-1.05\AA \ (35-50\% of the Ni--O distance), exceeding the Lindemann stability threshold (0.10--0.15\% bond length)~\cite{lindemann1984calculation}, so the lattice would likely destabilize before the resonance could be coherently reached.
Second, phonon damping competes with parametric gain, cutting the peak Raman response fivefold relative to the undamped case and collapsing the sustained spin-splitting plateau into a transient of only a few meV (\supf4). 
Third, the quartic IR anharmonicity ($\beta \approx 8\times 10^{-4}$) produces Duffing-type detuning of $Q_{\rm IR}$ from the fixed pump frequency as its amplitude grows, self-limiting the IR excitation, and capping the biquadratic term below the level needed to sustain parametric growth of $Q_\mathrm{R}$. Upon including this term, the Raman mode remains pinned near its initial amplitude regardless of pump strength (\supf5).
Taken together, these practical constraints suggest that, although NiO may not be ideal for applications, it clarifies where improvements are needed for the biquadratic-coupling mechanism---reduced damping, weaker IR anharmonicity, and a more favorable stability window. These features can also guide materials selection.

\paragraph*{Driven response in LaFeO$_3$.} 
We next apply the same framework to LaFeO$_3$ \cite{Scafetta2014}, a semiconducting antiferromagnet with equilibrium NRSS.
The orthorhombic perovskite hosts G-type order with magnetic space group $Pn'm'a$ and magnetic irrep $m\Gamma_2^+$, forming an altermagnetic state with NRSS that vanishes at $\Gamma$.
Our calculations confirm an equilibrium spin splitting $S=S_0=14.5$\,meV for the undistorted equilibrium structure (\supt1, \supf6).

Mode-selective excitations can access many more Raman-active modes in LaFeO$_3$ with NLPs and produce distinct dynamical responses.
A complete zone-center mode analysis, analogous to that presented for NiO (\autoref{tab:nio-lattice-dynamics}) together with the corresponding changes in $S$, is provided in \supt1 \cite{supp}. % of the SM \cite{supp}.
Here we focus on a representative excitation of a $\Gamma_{4}^+ (60)$ Raman mode (18.3 THz), which has the strongest modulation to $S$, to describe the two main effects. 
First, in the lossless model the existing NRSS is enhanced by $\sim$100\% ($\langle S\rangle_t = 28.3$ meV at 3\,ps) under a moderate pump field $E_0=1.5\,$MV/cm via a trilinear phonon coupling (\supt2). Including phonon damping, however, reduces the peak Raman amplitude by about an order of magnitude and collapses the enhancement to $\sim$6\% ($\langle S\rangle_t$ =15.4 \,meV at $t=1$\,ps, Fig.\ S9 \cite{supp}).
Second, excitation of the $\Gamma_{4}^+$ mode, consistent with the symmetry selection rules, induces a Zeeman-like spin splitting along $\Gamma$--$R$ that lifts the spin degeneracy at $\Gamma$ and generates a transient net moment (\supf6,7).
Both effects arise at modest Raman amplitudes of $|Q|=0.089$\,\AA$\sqrt{\mathrm{amu}}$, highlighting the efficiency of the phonon-driven symmetry breaking.

As in NiO, the induced moment in the presence of SOC enables direct optical detection of the dynamical NRSS.
In equilibrium, LaFeO$_3$ exhibits negligible Kerr response because its altermagnetic splitting is antisymmetric in momentum and yields no net magnetization. %
Under a $\Gamma_{4}^+$ Raman excitation, however, spin degeneracy is lifted, producing a finite moment detectable via polar MOKE. 
Polar MOKE measured in a geometry with the external field applied along the [010] direction, parallel to the spin axis of LaFeO$_{3}$, enables detection of the dynamically induced moment. 
Simultaneous excitation of the 8.6\,THz $\Gamma_2^- (33)$ IR mode and the 9.7\,THz $\Gamma_3^- (41)$ IR mode, which we find is strongly coupled (via trilinear coupling $\Gamma_2^-\otimes\Gamma_3^-=\Gamma_4^+$) to the 18.3\,THz $\Gamma_{4}^+ (60)$ Raman mode (\supt2), generates a time-dependent MOKE signal. %in polar geometry analogous to the NiO case. 
Simulations of the NLP dynamics and MOKE response are found in \supf8-11, where the buildup of the Raman coordinate tracks the emergence of both the Zeeman-like splitting and the Kerr rotation. 
A small static Kerr signal may arise from weak spin canting (weak ferromagnetic order from Dzyaloshinskii-Moriya exchange is symmetry allowed along the $c$ axis in undistorted LaFeO$_3$), but its modulation under resonant driving at the Raman mode frequency provides clear evidence of the nonequilibrium spin splitting. Complementary probes such as magnetic linear birefringence and dichroism, combined with uniaxial strain, offer additional sensitivity to the orbital-spin locking and multipolar order %of the altermagnetic state 
\cite{vila_orbital-spin_2025,sun_symmetry-breaking_2025,sunko_linear_2026}.

\paragraph{Experimental considerations.}
A key experimental issue to resolve is the role of antiferromagnetic domains. Because the driven phonon is spatially uniform, it does not select a domain orientation, and the measured MOKE signal reflects the pre-existing domain distribution. For example, scanning MOKE studies of the altermagnet MnTe reveal macroscopic domains with opposite Kerr rotation whose populations are set during cooling, while modest field cooling ($\sim$0.1\,T) can produce large single-domain regions \cite{Amin:2024}. Recent work on FeOF shows that nearly degenerate short-range chemically ordered clusters give rise to nanoscale domains with distinct NRSS characteristics and MOKE signatures, highlighting the sensitivity of domain structure to sample preparation \cite{Nathan:2026}.

Antiferromagnetic domains can be selectively enlarged in different magnetic materials by exploiting symmetry-allowed couplings to strain or external fields. For the specific materials considered here, several established routes exist. In NiO, oxygen annealing and strain engineering (bulk crystals and thin films) can increase domain sizes to well above typical pump spot diameters ($\sim$30--50\,$\mu$m).~\cite{PhysRevLett.105.077402,PhysRevB.74.144401} In LaFeO$_3$, epitaxial strain ~\cite{das_evolution_2018,zhu2024structural} and post-growth heat treatments ~\cite{polewczyk2025control} can similarly enlarge domains and bias domain populations, enabling a nonzero MOKE signal even under spatially uniform THz excitation.
Thus, domain detection using time-resolved MOKE microscopy with a probe spot below the characteristic domain size, or the use of domain-insensitive probes such as magnetic linear dichroism and magnetic second harmonic generation \cite{vila_orbital-spin_2025,ma_probing_2025,jeong_altermagnetic_2026}, provide practical routes to avoid cancellation from domain averaging.

\paragraph{Conclusion.} 
We have demonstrated that nonlinear phononics enables ultrafast dynamical control of nonrelativistic spin splitting in collinear antiferromagnets, extending beyond the reach of static strain. 
The two symmetry criteria derived here, wavevector compatibility with the magnetic propagation vector and phonon-magnetic inversion-parity matching, reduce the identification of dynamical NRSS-activating modes to a systematic group-theoretical process applicable to any
collinear antiferromagnet, replacing prior heuristic approaches. 
We used NiO and LaFeO$_3$ as representative cases to illustrate two distinct anharmonic pathways consistent with our symmetry selection rules: a biquadratic (quartic) IR--Raman coupling that generate NRSS from a spin-degenerate ground state (NiO), and a trilinear (cubic) coupling that drives an NRSS-active Raman distortion and amplifies an intrinsic altermagnetic splitting (LaFeO$_3$).
Rectifying couplings of the form $Q_\mathrm{R}Q_\mathrm{IR}^2$ can prolong symmetry lowering, but they typically involve an $A_{1g}$-like ($\Gamma_1^+$) Raman coordinate that is not itself NRSS-active; in MnPS$_3$ the reported splitting \cite{PhysRevB.108.054308} is instead primarily induced by the resonantly driven IR mode.
Additional scenarios, including the dynamical generation of antisymmetric (odd-parity) NRSS, are symmetry-allowed within the same framework and remain to be explored.
    We note that optical control of crystal structure and magnetism via selective phonon driving has been explored in prior NLP and ultrafast materials-engineering studies~\cite{disa_engineering_2021, gu_ultrafast_2016,PhysRevB.108.054308,c3kh-5s5t}. Beyond demonstrating a route to optical lattice control, we establish a predictive symmetry framework for dynamical NRSS and identify candidates for experimental realization.
These results establish nonlinear phononics as a general, symmetry-guided design for controlling spin splitting in antiferromagnets on ultrafast timescales, extending dynamical lattice control into the manipulation of momentum-resolved spin textures with speeds that static approaches cannot achieve.

%Reference~\cite{rajpurohit2026symmetry} reports phonon-induced NRSS in MnPS$_3$, whose magnetic order transforms as $m\Gamma_2^-$. Our wavevector-compatibility and parity-matching criteria therefore require an NRSS-activating mode at $\Gamma$ with odd inversion parity, consistent with the infrared-active $\Gamma$ mode identified in that work. Reference~\cite{wang2026ultrafast} investigates the dynamical lifting of spin degeneracy at $\Gamma$, analogous to our LaFeO$_3$ example. Together, these studies complement our findings and illustrate the applicability of our mode-selection rules beyond the two materials considered here.}

\begin{acknowledgments}
Work at Northwestern University was supported by the Air Force Office of Scientific Research (AFOSR) under award number FA9550-23-1-0042. Work at Cornell University was supported by the U.S.\ Department of Energy, Office of Science, Office of Basic Energy Sciences, under Contract No.\ DE-SC0019414. L.Y.\  acknowledges the use of Claude Opus4.8 for generating part of the source code used to analysis multi frozen phonon modes energy landscape and simulate phonon dynamics. Generated code were reviewed manually and tested against various parameter choices, results are consistent with symmetry rules and physical intuitions of nonlinearly coupled oscillators. Experimental considerations were informed, in part, by discussions at the Aspen Center for Physics, which is supported by the  National Science Foundation (NSF) grant PHY-2210452. 
All first-principles calculations and analyses were conducted using the computational resources at the Quest high-performance computing facility at Northwestern University, which is jointly supported by the Office of the Provost, the Office for Research, and Northwestern University Information Technology.

After submitting this work, we became aware of two related preprints on dynamical NRSS \cite{rajpurohit2026symmetry,wang2026ultrafast}. These studies explore complementary, material-specific realizations of the phenomena consistent with our symmetry framework.
\end{acknowledgments}

\bibliographystyle{apsrev4-2}
\bibliography{references}

@article{zunger2018inverse,
  title={Inverse design in search of materials with target functionalities},
  author={Zunger, Alex},
  journal={Nature Reviews Chemistry},
  volume={2},
  number={4},
  pages={0121},
  year={2018},
  doi = {10.1038/s41570-018-0121},
  publisher={Nature Publishing Group UK London}
}

@article{c3kh-5s5t,
  title = {Rational control of magnonic and electronic band splittings},
  author = {Bandyopadhyay, Subhadeep and Raj, Anoop and Ghosez, Philippe and Pujari, Sumiran and Bhowal, Sayantika},
  journal = {Phys. Rev. B},
  volume = {112},
  issue = {10},
  pages = {L100402},
  numpages = {7},
  year = {2025},
  month = {Sep},
  publisher = {American Physical Society},
  doi = {10.1103/c3kh-5s5t},
  url = {https://link.aps.org/doi/10.1103/c3kh-5s5t}
}

@article{forst_nonlinear_2011,
        title = {Nonlinear phononics as an ultrafast route to lattice control},
        volume = {7},
        issn = {1745-2481},
        url = {https://doi.org/10.1038/nphys2055},
        doi = {10.1038/nphys2055},
        number = {11},
        journal = {Nature Physics},
        author = {Först, M. and Manzoni, C. and Kaiser, S. and Tomioka, Y. and Tokura, Y. and Merlin, R. and Cavalleri, A.},
        month = nov,
        year = {2011},
        pages = {854--856},
}

@article{fausti2011light,
  title={Light-induced superconductivity in a stripe-ordered cuprate},
  author={Fausti, Daniele and Tobey, RI and Dean, Nicky and Kaiser, Stefan and Dienst, A and Hoffmann, Matthias C and Pyon, S and Takayama, T and Takagi, H and Cavalleri, Andrea},
  journal={Science},
  volume={331},
  number={6014},
  pages={189--191},
  year={2011},
  doi={10.1126/science.1197294},
  publisher={American Association for the Advancement of Science}
}

@article{PhysRevB.97.085145,
  title = {Breaking symmetry with light: Ultrafast ferroelectricity and magnetism from three-phonon coupling},
  author = {Radaelli, Paolo G.},
  journal = {Phys. Rev. B},
  volume = {97},
  issue = {8},
  pages = {085145},
  numpages = {9},
  year = {2018},
  month = {Feb},
  publisher = {American Physical Society},
  doi = {10.1103/PhysRevB.97.085145},
  url = {https://link.aps.org/doi/10.1103/PhysRevB.97.085145}
}

@article{noda2016momentum,
  title={Momentum-dependent band spin splitting in semiconducting MnO 2: a density functional calculation},
  author={Noda, Yusuke and Ohno, Kaoru and Nakamura, Shinichiro},
  journal={Physical Chemistry Chemical Physics},
  volume={18},
  number={19},
  pages={13294--13303},
  year={2016},
  doi = {10.1039/c5cp07806g},
  publisher={Royal Society of Chemistry}
}

@article{PhysRevB.102.014422,
  title = {{Giant momentum-dependent spin splitting in centrosymmetric low-$Z$ antiferromagnets}},
  author = {Yuan, Lin-Ding and Wang, Zhi and Luo, Jun-Wei and Rashba, Emmanuel I. and Zunger, Alex},
  journal = {Phys. Rev. B},
  volume = {102},
  issue = {1},
  pages = {014422},
  numpages = {13},
  year = {2020},
  month = {Jul},
  publisher = {American Physical Society},
  doi = {10.1103/PhysRevB.102.014422},
  url = {https://link.aps.org/doi/10.1103/PhysRevB.102.014422}
}

@article{PhysRevMaterials.5.014409,
  title = {{Prediction of low-Z collinear and noncollinear antiferromagnetic compounds having momentum-dependent spin splitting even without spin-orbit coupling}},
  author = {Yuan, Lin-Ding and Wang, Zhi and Luo, Jun-Wei and Zunger, Alex},
  journal = {Phys. Rev. Mater.},
  volume = {5},
  issue = {1},
  pages = {014409},
  numpages = {24},
  year = {2021},
  month = {Jan},
  publisher = {American Physical Society},
  doi = {10.1103/PhysRevMaterials.5.014409},
  url = {https://link.aps.org/doi/10.1103/PhysRevMaterials.5.014409}
}

@article{https://doi.org/10.1002/adma.202211966,
author = {Yuan, Lin-Ding and Zunger, Alex},
title = {{Degeneracy Removal of Spin Bands in Collinear Antiferromagnets with Non-Interconvertible Spin-Structure Motif Pair}},
journal = {Advanced Materials},
volume = {35},
number = {31},
pages = {2211966},
doi={10.1002/adma.202211966},
doi = {https://doi.org/10.1002/adma.202211966},
url = {https://onlinelibrary.wiley.com/doi/abs/10.1002/adma.202211966},
year = {2023}
}

@article{yuan2023degeneracy,
  title={Degeneracy Removal of Spin Bands in Collinear Antiferromagnets with Non-Interconvertible Spin-Structure Motif Pair},
  author={Yuan, Lin-Ding and Zunger, Alex},
  journal={Advanced Materials},
  volume={35},
  number={31},
  pages={2211966},
  year={2023},
  publisher={Wiley Online Library}
}

@article{PhysRevB.99.184432,
  title = {{Antiferromagnetism in ${\mathrm{RuO}}_{2}$ as $d$-wave Pomeranchuk instability}},
  author = {Ahn, Kyo-Hoon and Hariki, Atsushi and Lee, Kwan-Woo and Kune\ifmmode \check{s}\else \v{s}\fi{}, Jan},
  journal = {Phys. Rev. B},
  volume = {99},
  issue = {18},
  pages = {184432},
  numpages = {5},
  year = {2019},
  month = {May},
  publisher = {American Physical Society},
  doi = {10.1103/PhysRevB.99.184432},
  url = {https://link.aps.org/doi/10.1103/PhysRevB.99.184432}
}

@Article{Naka2019,
author={Naka, Makoto
and Hayami, Satoru
and Kusunose, Hiroaki
and Yanagi, Yuki
and Motome, Yukitoshi
and Seo, Hitoshi},
title={{Spin current generation in organic antiferromagnets}},
journal={Nature Communications},
year={2019},
month={Sep},
day={20},
volume={10},
number={1},
pages={4305},
issn={2041-1723},
doi={10.1038/s41467-019-12229-y},
url={https://doi.org/10.1038/s41467-019-12229-y}
}

@article{PhysRevB.101.220403,
  title = {{Spontaneous antisymmetric spin splitting in noncollinear antiferromagnets without spin-orbit coupling}},
  author = {Hayami, Satoru and Yanagi, Yuki and Kusunose, Hiroaki},
  journal = {Phys. Rev. B},
  volume = {101},
  issue = {22},
  pages = {220403},
  numpages = {6},
  year = {2020},
  month = {Jun},
  publisher = {American Physical Society},
  doi = {10.1103/PhysRevB.101.220403},
  url = {https://link.aps.org/doi/10.1103/PhysRevB.101.220403}
}

@article{doi:10.7566/JPSJ.88.123702,
author = {Hayami ,Satoru and Yanagi ,Yuki and Kusunose ,Hiroaki},
title = {{Momentum-Dependent Spin Splitting by Collinear Antiferromagnetic Ordering}},
journal = {Journal of the Physical Society of Japan},
volume = {88},
number = {12},
pages = {123702},
year = {2019},
doi = {10.7566/JPSJ.88.123702},
URL = { 
        https://doi.org/10.7566/JPSJ.88.123702
}
}

@article{PhysRevB.102.144441,
  title = {{Bottom-up design of spin-split and reshaped electronic band structures in antiferromagnets without spin-orbit coupling: Procedure on the basis of augmented multipoles}},
  author = {Hayami, Satoru and Yanagi, Yuki and Kusunose, Hiroaki},
  journal = {Phys. Rev. B},
  volume = {102},
  issue = {14},
  pages = {144441},
  numpages = {24},
  year = {2020},
  month = {Oct},
  publisher = {American Physical Society},
  doi = {10.1103/PhysRevB.102.144441},
  url = {https://link.aps.org/doi/10.1103/PhysRevB.102.144441}
}

@article{
doi:10.1126/sciadv.aaz8809,
author = {Libor Šmejkal  and Rafael González-Hernández  and T. Jungwirth  and J. Sinova },
title = {{Crystal time-reversal symmetry breaking and spontaneous Hall effect in collinear antiferromagnets}},
journal = {Science Advances},
volume = {6},
number = {23},
pages = {eaaz8809},
year = {2020},
doi = {10.1126/sciadv.aaz8809},
URL = {https://www.science.org/doi/abs/10.1126/sciadv.aaz8809}}

@article{ma2021multifunctional,
  title={Multifunctional antiferromagnetic materials with giant piezomagnetism and noncollinear spin current},
  author={Ma, Hai-Yang and Hu, Mengli and Li, Nana and Liu, Jianpeng and Yao, Wang and Jia, Jin-Feng and Liu, Junwei},
  journal={Nature Communications},
  volume={12},
  number={1},
  pages={2846},
  year={2021},
  doi={10.1038/s41467-021-23127-7},
  publisher={Nature Publishing Group UK London}
}

@article{fender2025altermagnetism,
  title={Altermagnetism: A chemical perspective},
  author={Fender, Shannon S and Gonzalez, Oscar and Bediako, D Kwabena},
  journal={Journal of the American Chemical Society},
  volume={147},
  number={3},
  pages={2257--2274},
  year={2025},
  doi={10.1021/jacs.4c14503},
  publisher={ACS Publications}
}

@article{song2025altermagnets,
  title={Altermagnets as a new class of functional materials},
  author={Song, Cheng and Bai, Hua and Zhou, Zhiyuan and Han, Lei and Reichlova, Helena and Dil, J Hugo and Liu, Junwei and Chen, Xianzhe and Pan, Feng},
  journal={Nature Reviews Materials},
  volume={10},
  number={6},
  pages={473--485},
  year={2025},
  doi={10.1038/s41578-025-00779-1},
  publisher={Nature Publishing Group UK London}
}

@article{PhysRevX.12.031042,
  title = {{Beyond Conventional Ferromagnetism and Antiferromagnetism: A Phase with Nonrelativistic Spin and Crystal Rotation Symmetry}},
  author = {\ifmmode \check{S}\else \v{S}\fi{}mejkal, Libor and Sinova, Jairo and Jungwirth, Tomas},
  journal = {Phys. Rev. X},
  volume = {12},
  issue = {3},
  pages = {031042},
  numpages = {16},
  year = {2022},
  month = {Sep},
  publisher = {American Physical Society},
  doi = {10.1103/PhysRevX.12.031042},
  url = {https://link.aps.org/doi/10.1103/PhysRevX.12.031042}
}

@article{gu_ultrafast_2016,
        title = {Ultrafast {Band} {Engineering} and {Transient} {Spin} {Currents} in {Antiferromagnetic} {Oxides}},
        volume = {6},
        issn = {2045-2322},
        url = {https://doi.org/10.1038/srep25121},
        doi = {10.1038/srep25121},
        number = {1},
        journal = {Scientific Reports},
        author = {Gu, Mingqiang and Rondinelli, James M.},
        month = apr,
        year = {2016},
        pages = {25121},
}

@article{disa_polarizing_2020,
        title = {Polarizing an antiferromagnet by optical engineering of the crystal field},
        volume = {16},
        issn = {1745-2481},
        url = {https://doi.org/10.1038/s41567-020-0936-3},
        doi = {10.1038/s41567-020-0936-3},
        number = {9},
        journal = {Nature Physics},
        author = {Disa, Ankit S. and Fechner, Michael and Nova, Tobia F. and Liu, Biaolong and Först, Michael and Prabhakaran, Dharmalingam and Radaelli, Paolo G. and Cavalleri, Andrea},
        month = sep,
        year = {2020},
        pages = {937--941},
}

@article{zhou_manipulation_2025,
        title = {Manipulation of the altermagnetic order in {CrSb} via crystal symmetry},
        volume = {638},
        issn = {1476-4687},
        url = {https://doi.org/10.1038/s41586-024-08436-3},
        doi = {10.1038/s41586-024-08436-3},
        number = {8051},
        journal = {Nature},
        author = {Zhou, Zhiyuan and Cheng, Xingkai and Hu, Mengli and Chu, Ruiyue and Bai, Hua and Han, Lei and Liu, Junwei and Pan, Feng and Song, Cheng},
        month = feb,
        year = {2025},
        pages = {645--650},
}

@incollection{opechowski1965magnetic,
  author       = {Opechowski, W. and Guccione, R.},
  title        = {Magnetic Symmetry},
  booktitle    = {Magnetism},
  editor       = {Rado, G. T. and Suhl, H.},
  volume       = {2A},
  publisher    = {Academic Press},
  year         = {1965}
}

@article{bertaut1968representation,
  author       = {Bertaut, E. F.},
  title        = {Representation Analysis of Magnetic Structures},
  journal      = {Acta Crystallographica Section A},
  volume       = {24},
  pages        = {217--231},
  year         = {1968},
  doi={10.1107/S0567739468000306}
}

@article{shull1951neutron,
  title={Neutron diffraction by paramagnetic and antiferromagnetic substances},
  author={Shull, C Gi and Strauser, WA and Wollan, EO},
  journal={Physical Review},
  volume={83},
  number={2},
  pages={333},
  year={1951},
  doi={10.1103/PhysRev.83.333},
  publisher={APS}
}

@article{baruchel1981antiferromagnetic,
  title={Antiferromagnetic S-domains in NiO: I. Neutron magnetic topographic investigation},
  author={Baruchel, J and Schlenker, M and Kurosawa, K and Saito, S},
  journal={Philosophical Magazine B},
  volume={43},
  number={5},
  pages={853--860},
  year={1981},
  doi={10.1080/01418638108222351},
  publisher={Taylor \& Francis}
}

@article{subedi_theory_2014,
    title = {Theory of nonlinear phononics for coherent light control of solids},
    volume = {89},
    copyright = {http://link.aps.org/licenses/aps-default-license},
    issn = {1098-0121, 1550-235X},
    url = {https://link.aps.org/doi/10.1103/PhysRevB.89.220301},
    doi = {10.1103/PhysRevB.89.220301},
    language = {en},
    number = {22},
    urldate = {2025-10-05},
    journal = {Physical Review B},
    author = {Subedi, Alaska and Cavalleri, Andrea and Georges, Antoine},
    month = jun,
    year = {2014},
    pages = {220301},
}

@article{vila_orbital-spin_2025,
	title = {Orbital-spin locking and its optical signatures in altermagnets},
	volume = {112},
	issn = {2469-9950, 2469-9969},
	url = {https://link.aps.org/doi/10.1103/bzzy-ngcs},
	doi = {10.1103/bzzy-ngcs},
	language = {en},
	number = {2},
	urldate = {2026-06-08},
	journal = {Physical Review B},
	author = {Vila, Marc and Sunko, Veronika and Moore, Joel E.},
	month = jul,
	year = {2025},
	pages = {L020401},
}

@article{sun_symmetry-breaking_2025,
	title = {Symmetry-{Breaking} {Magneto}-{Optical} {Effects} in {Altermagnets}},
	volume = {25},
	copyright = {https://doi.org/10.15223/policy-029},
	issn = {1530-6984, 1530-6992},
	url = {https://pubs.acs.org/doi/10.1021/acs.nanolett.5c03647},
	doi = {10.1021/acs.nanolett.5c03647},
	language = {en},
	number = {41},
	urldate = {2026-06-08},
	journal = {Nano Letters},
	author = {Sun, Jiuyu and Du, Yongping and Kan, Erjun},
	month = oct,
	year = {2025},
	pages = {14960--14966},
}

@article{sunko_linear_2026,
	title = {Linear magneto-birefringence as a probe of altermagnetism},
	copyright = {2026 The Author(s)},
	issn = {2397-4648},
	url = {https://www.nature.com/articles/s41535-026-00901-8},
	doi = {10.1038/s41535-026-00901-8},
	language = {en},
	urldate = {2026-06-08},
	journal = {npj Quantum Materials},
	author = {Sunko, V. and Orenstein, J.},
	month = may,
	year = {2026},
}

@article{ma_probing_2025,
	title = {Probing spin-split bands in altermagnets through second harmonic generation},
	volume = {111},
	issn = {2469-9950, 2469-9969},
	url = {https://link.aps.org/doi/10.1103/PhysRevB.111.064311},
	doi = {10.1103/PhysRevB.111.064311},
	language = {en},
	number = {6},
	urldate = {2026-06-15},
	journal = {Physical Review B},
	author = {Ma, Y. X. and Chang, J. Z. and Liu, Y. D. and Si, M. S. and Zhang, G. P. and Zhang, Z. M.},
	month = feb,
	year = {2025},
	pages = {064311},
}

@article{jeong_altermagnetic_2026,
	title = {Altermagnetic polar metallic phase in ultrathin epitaxially strained {RuO}$_{\textrm{2}}$ films},
	volume = {123},
	issn = {0027-8424, 1091-6490},
	url = {https://pnas.org/doi/10.1073/pnas.2526641123},
	doi = {10.1073/pnas.2526641123},
	language = {en},
	number = {10},
	urldate = {2026-06-15},
	journal = {Proceedings of the National Academy of Sciences},
	author = {Jeong, Seung Gyo and Choi, In Hyeok and Nair, Sreejith and Buiarelli, Luca and Pourbahari, Bita and Oh, Jin Young and Lin, Bonnie Y.X. and LeBeau, James M. and Bassim, Nabil and Hirai, Daigorou and Seo, Ambrose and Choi, Woo Seok and Fernandes, Rafael M. and Birol, Turan and Zhao, Liuyan and Lee, Jong Seok and Jalan, Bharat},
	month = mar,
	year = {2026},
	pages = {e2526641123},
}

@article{10.1063/1.4928289,
    author = {Rondinelli, James M. and Poeppelmeier, Kenneth R. and Zunger, Alex},
    title = {Research Update: Towards designed functionalities in oxide-based electronic materials},
    journal = {APL Materials},
    volume = {3},
    number = {8},
    pages = {080702},
    year = {2015},
    month = {08},
        doi = {10.1063/1.4928289},
    url = {https://doi.org/10.1063/1.4928289},
}

@misc{supp,
note = {See Supplemental Material at [URL will be inserted by publisher] for details of the first-principles calculations, lattice-dynamical and nonlinear-phononics modeling, symmetry analysis establishing the criteria for phonon-induced nonrelativistic spin splitting, derivation of phonon-coupling selection rules, additional results for NiO and LaFeO$_3$, including electronic structures, pump-field and damping dependences, Kerr-rotation spectra, spin-canting and magnetization analyses, and fitted anharmonic coupling parameters and phonon-dynamics simulations.}
}

@article{Scafetta2014,
  title = {Band structure and optical transitions in LaFeO<sub>3</sub>: theory and experiment},
  volume = {26},
  ISSN = {1361-648X},
  url = {http://dx.doi.org/10.1088/0953-8984/26/50/505502},
  DOI = {10.1088/0953-8984/26/50/505502},
  number = {50},
  journal = {Journal of Physics: Condensed Matter},
  publisher = {IOP Publishing},
  author = {Scafetta,  Mark D and Cordi,  Adam M and Rondinelli,  James M and May,  Steven J},
  year = {2014},
  month = Nov,
  pages = {505502}
}

@article{Nathan:2026,
  title = {Anion correlation induced nonrelativistic spin splitting in rutile antiferromagnets},
  author = {Nathan, Siddhartha S. and Puggioni, Danilo and Yuan, Linding and Rondinelli, James M.},
  journal = {Phys. Rev. Mater.},
  volume = {10},
  issue = {5},
  pages = {054404},
  numpages = {9},
  year = {2026},
  month = {May},
  publisher = {American Physical Society},
  doi = {10.1103/vd4d-cxml},
  url = {https://link.aps.org/doi/10.1103/vd4d-cxml}
}

@article{Amin:2024,
  title = {Nanoscale imaging and control of altermagnetism in MnTe},
  volume = {636},
  ISSN = {1476-4687},
  url = {http://dx.doi.org/10.1038/s41586-024-08234-x},
  DOI = {10.1038/s41586-024-08234-x},
  number = {8042},
  journal = {Nature},
  publisher = {Springer Science and Business Media LLC},
  author = {Amin,  O. J. and Dal Din,  A. and Golias,  E. and Niu,  Y. and Zakharov,  A. and Fromage,  S. C. and Fields,  C. J. B. and Heywood,  S. L. and Cousins,  R. B. and Maccherozzi,  F. and Krempaský,  J. and Dil,  J. H. and Kriegner,  D. and Kiraly,  B. and Campion,  R. P. and Rushforth,  A. W. and Edmonds,  K. W. and Dhesi,  S. S. and Šmejkal,  L. and Jungwirth,  T. and Wadley,  P.},
  year = {2024},
  month = Dec,
  pages = {348–353}
}

@techreport{lindemann1984calculation,
  title={The calculation of molecular Eigen-frequencies},
  author={Lindemann, Frederick A},
  year={1910},
  journal = {Physikalische Zeitschrift}, 
  volume = {11}, 
  pages = {609–612}
}

@article{1r6k-s46h,
  title = {Designing nonrelativistic spin splitting in oxide perovskites},
  author = {Bandyopadhyay, Subhadeep and Picozzi, Silvia and Bhowal, Sayantika},
  journal = {Phys. Rev. B},
  volume = {112},
  issue = {6},
  pages = {064405},
  numpages = {15},
  year = {2025},
  month = {Aug},
  publisher = {American Physical Society},
  doi = {10.1103/1r6k-s46h},
  url = {https://link.aps.org/doi/10.1103/1r6k-s46h}
}

@article{jungwirth_altermagnetic_2026,
	title = {Altermagnetic spintronics},
	issn = {1745-2481},
	url = {https://doi.org/10.1038/s41567-026-03337-w},
	doi = {10.1038/s41567-026-03337-w},
	journal = {Nature Physics},
	author = {Jungwirth, T. and Sinova, J. and Wadley, P. and Kriegner, D. and Reichlová, H. and Krizek, F. and Ohno, H. and Šmejkal, L.},
	month = jul,
	year = {2026},
}

@article{mankowsky_nonlinear_2014,
	title = {Nonlinear lattice dynamics as a basis for enhanced superconductivity in {YBa2Cu3O6}.5},
	volume = {516},
	issn = {1476-4687},
	url = {https://doi.org/10.1038/nature13875},
	doi = {10.1038/nature13875},
	number = {7529},
	journal = {Nature},
	author = {Mankowsky, R. and Subedi, A. and Först, M. and Mariager, S. O. and Chollet, M. and Lemke, H. T. and Robinson, J. S. and Glownia, J. M. and Minitti, M. P. and Frano, A. and Fechner, M. and Spaldin, N. A. and Loew, T. and Keimer, B. and Georges, A. and Cavalleri, A.},
	month = dec,
	year = {2014},
	pages = {71--73},
}

@article{davies_phononic_2024,
	title = {Phononic switching of magnetization by the ultrafast {Barnett} effect},
	volume = {628},
	issn = {1476-4687},
	url = {https://doi.org/10.1038/s41586-024-07200-x},
	doi = {10.1038/s41586-024-07200-x},
	number = {8008},
	journal = {Nature},
	author = {Davies, C. S. and Fennema, F. G. N. and Tsukamoto, A. and Razdolski, I. and Kimel, A. V. and Kirilyuk, A.},
	month = apr,
	year = {2024},
	pages = {540--544},
}

@article{stupakiewicz_ultrafast_2021,
	title = {Ultrafast phononic switching of magnetization},
	volume = {17},
	issn = {1745-2481},
	url = {https://doi.org/10.1038/s41567-020-01124-9},
	doi = {10.1038/s41567-020-01124-9},
	number = {4},
	journal = {Nature Physics},
	author = {Stupakiewicz, A. and Davies, C. S. and Szerenos, K. and Afanasiev, D. and Rabinovich, K. S. and Boris, A. V. and Caviglia, A. and Kimel, A. V. and Kirilyuk, A.},
	month = apr,
	year = {2021},
	pages = {489--492},
}

@article{disa_photo-induced_2023,
	title = {Photo-induced high-temperature ferromagnetism in {YTiO3}},
	volume = {617},
	issn = {1476-4687},
	url = {https://doi.org/10.1038/s41586-023-05853-8},
	doi = {10.1038/s41586-023-05853-8},
	number = {7959},
	journal = {Nature},
	author = {Disa, A. S. and Curtis, J. and Fechner, M. and Liu, A. and von Hoegen, A. and Först, M. and Nova, T. F. and Narang, P. and Maljuk, A. and Boris, A. V. and Keimer, B. and Cavalleri, A.},
	month = may,
	year = {2023},
	pages = {73--78},
}

@article{
doi:10.1126/science.adi9601,
author = {Jiaming Luo  and Tong Lin  and Junjie Zhang  and Xiaotong Chen  and Elizabeth R. Blackert  and Rui Xu  and Boris I. Yakobson  and Hanyu Zhu },
title = {Large effective magnetic fields from chiral phonons in rare-earth halides},
journal = {Science},
volume = {382},
number = {6671},
pages = {698-702},
year = {2023},
doi = {10.1126/science.adi9601},
URL = {https://www.science.org/doi/abs/10.1126/science.adi9601},
eprint = {https://www.science.org/doi/pdf/10.1126/science.adi9601}}

@article{PhysRevLett.118.197601,
  title = {Ultrafast Reversal of the Ferroelectric Polarization},
  author = {Mankowsky, R. and von Hoegen, A. and F\"orst, M. and Cavalleri, A.},
  journal = {Phys. Rev. Lett.},
  volume = {118},
  issue = {19},
  pages = {197601},
  numpages = {5},
  year = {2017},
  month = {May},
  publisher = {American Physical Society},
  doi = {10.1103/PhysRevLett.118.197601},
  url = {https://link.aps.org/doi/10.1103/PhysRevLett.118.197601}
}

@article{
doi:10.1126/science.adr4713,
author = {Z. Zeng  and M. Först  and M. Fechner  and M. Buzzi  and E. B. Amuah  and C. Putzke  and P. J. W. Moll  and D. Prabhakaran  and P. G. Radaelli  and A. Cavalleri },
title = {Photo-induced chirality in a nonchiral crystal},
journal = {Science},
volume = {387},
number = {6732},
pages = {431-436},
year = {2025},
doi = {10.1126/science.adr4713},
URL = {https://www.science.org/doi/abs/10.1126/science.adr4713},
eprint = {https://www.science.org/doi/pdf/10.1126/science.adr4713}}

@article{
doi:10.1126/science.adz5230,
author = {Z. Zeng  and M. Först  and M. Fechner  and D. Prabhakaran  and P. G. Radaelli  and A. Cavalleri },
title = {Photo-induced nonvolatile rewritable ferroaxial switching},
journal = {Science},
volume = {390},
number = {6769},
pages = {195-198},
year = {2025},
doi = {10.1126/science.adz5230},
URL = {https://www.science.org/doi/abs/10.1126/science.adz5230},
eprint = {https://www.science.org/doi/pdf/10.1126/science.adz5230}}

@article{PhysRevMaterials.2.064401,
  title = {Magnetophononics: Ultrafast spin control through the lattice},
  author = {Fechner, M. and Sukhov, A. and Chotorlishvili, L. and Kenel, C. and Berakdar, J. and Spaldin, N. A.},
  journal = {Phys. Rev. Mater.},
  volume = {2},
  issue = {6},
  pages = {064401},
  numpages = {11},
  year = {2018},
  month = {Jun},
  publisher = {American Physical Society},
  doi = {10.1103/PhysRevMaterials.2.064401},
  url = {https://link.aps.org/doi/10.1103/PhysRevMaterials.2.064401}
}

@article{
doi:10.1126/sciadv.ado0722,
author = {Tom Kahana  and Daniel A. Bustamante Lopez  and Dominik M. Juraschek },
title = {Light-induced magnetization from magnonic rectification},
journal = {Science Advances},
volume = {10},
number = {39},
pages = {eado0722},
year = {2024},
doi = {10.1126/sciadv.ado0722},
URL = {https://www.science.org/doi/abs/10.1126/sciadv.ado0722},
eprint = {https://www.science.org/doi/pdf/10.1126/sciadv.ado0722}}

@article{khalsa_ultrafast_2018,
	title = {Ultrafast optically induced ferromagnetic/anti-ferromagnetic phase transition in {GdTiO3} from first principles},
	volume = {3},
	issn = {2397-4648},
	url = {https://doi.org/10.1038/s41535-018-0086-3},
	doi = {10.1038/s41535-018-0086-3},
	number = {1},
	journal = {npj Quantum Materials},
	author = {Khalsa, Guru and Benedek, Nicole A.},
	month = mar,
	year = {2018},
	pages = {15},
}

@article{ismail-beigi_picoscale_2017,
	title = {Picoscale materials engineering},
	volume = {2},
	issn = {2058-8437},
	url = {https://doi.org/10.1038/natrevmats.2017.60},
	doi = {10.1038/natrevmats.2017.60},
	number = {11},
	journal = {Nature Reviews Materials},
	author = {Ismail-Beigi, Sohrab and Walker, Frederick J. and Disa, Ankit S. and Rabe, Karin M. and Ahn, Charles H.},
	month = sep,
	year = {2017},
	pages = {17060},
}

@article{disa_engineering_2021,
	title = {Engineering crystal structures with light},
	volume = {17},
	issn = {1745-2481},
	url = {https://doi.org/10.1038/s41567-021-01366-1},
	doi = {10.1038/s41567-021-01366-1},
	number = {10},
	journal = {Nature Physics},
	author = {Disa, Ankit S. and Nova, Tobia F. and Cavalleri, Andrea},
	month = oct,
	year = {2021},
	pages = {1087--1092},
}

@article{PhysRevLett.118.054101,
  title = {Ultrafast Structure Switching through Nonlinear Phononics},
  author = {Juraschek, D. M. and Fechner, M. and Spaldin, N. A.},
  journal = {Phys. Rev. Lett.},
  volume = {118},
  issue = {5},
  pages = {054101},
  numpages = {5},
  year = {2017},
  month = {Jan},
  publisher = {American Physical Society},
  doi = {10.1103/PhysRevLett.118.054101},
  url = {https://link.aps.org/doi/10.1103/PhysRevLett.118.054101}
}

@article{PhysRevB.108.054308,
  title = {Nonlinear phononic coupling in antiferromagnetic monolayer ${\mathrm{MnPS}}_{3}$},
  author = {Zeng, Yijie},
  journal = {Phys. Rev. B},
  volume = {108},
  issue = {5},
  pages = {054308},
  numpages = {8},
  year = {2023},
  month = {Aug},
  publisher = {American Physical Society},
  doi = {10.1103/PhysRevB.108.054308},
  url = {https://link.aps.org/doi/10.1103/PhysRevB.108.054308}
}

@article{PhysRevB.109.024110,
  title = {Coherent control of the translational and point group symmetries of crystals with light},
  author = {Khalsa, Guru and Kaaret, Jeffrey Z. and Benedek, Nicole A.},
  journal = {Phys. Rev. B},
  volume = {109},
  issue = {2},
  pages = {024110},
  numpages = {15},
  year = {2024},
  month = {Jan},
  publisher = {American Physical Society},
  doi = {10.1103/PhysRevB.109.024110},
  url = {https://link.aps.org/doi/10.1103/PhysRevB.109.024110}
}

@article{rajpurohit2026symmetry,
  title={Symmetry-selective nonrelativistic spin splitting in antiferromagnets driven by coherent phonons},
  author={Rajpurohit, Sangeeta and Yarmohammadi, Mohsen and Haque, Sheikh Rubaiat Ul and Heinz, Tony F and Lindenberg, Aaron M and Ogitsu, Tadashi},
  journal={arXiv preprint arXiv:2609.01844},
  year={2026}
}

@article{wang2026ultrafast,
  title={Ultrafast altermagnetophononics},
  author={Wang, Chenyu and Wang, Yaxian and Meng, Sheng},
  journal={arXiv preprint arXiv:2607.13863},
  year={2026}
}

@article{lu_nonlinear_2024,
  title        = {Nonlinear optical physics at terahertz frequency},
  author       = {Lu, Yao and Huang, Yibo and Cheng, Junkai and Ma, Ruobin and Xu, Xitan and Zang, Yijia and Wu, Qiang and Xu, Jingjun},
  journal      = {Nanophotonics},
  year         = {2024},
  volume       = {13},
  number       = {18},
  pages        = {3279--3294},
  doi          = {10.1515/nanoph-2024-0109}
}

@article{lu_stimulated_2022,
  title        = {Light--matter interaction beyond Born--Oppenheimer approximation mediated by stimulated phonon polaritons},
  author       = {Lu, Yao and Wu, Qiang and Xiong, Hao and Zhou, Xu and Li, Zhixuan and Qi, Jiwei and Xu, Xitan and Ma, Ruobin and Fan, Jiang and Chen, Zhigang and Xu, Jingjun},
  journal      = {Communications Physics},
  year         = {2022},
  volume       = {5},
  pages        = {299},
  doi          = {10.1038/s42005-022-01080-1}
}

@article{zhang_tunable_2026,
  title = {Tunable {Hidden} {Altermagnetic} {Spin} {Splitting} in {Layered} {Ruddlesden}–{Popper} {Oxides}},
  volume = {26},
  copyright = {https://doi.org/10.15223/policy-029},
  issn = {1530-6984, 1530-6992},
  url = {https://pubs.acs.org/doi/10.1021/acs.nanolett.6c00013},
  doi = {10.1021/acs.nanolett.6c00013},
  language = {en},
  number = {7},
  urldate = {2026-05-24},
  journal = {Nano Letters},
  author = {Zhang, Tongxie and Yuan, Linding and Rondinelli, James M. and Fertig, H. A. and Zhang, Shixiong},
  month = feb,
  year = {2026},
  pages = {2778--2786},
}

@article{han2026strain,
  title={Strain-Tunable Altermagnetism in Bilayer Perovskites Mediated by Jahn--Teller Distortions},
  author={Han, Peng and Zhu, Rui and Huang, Hanle and Cheng, Zhengguang and Zhang, Jingtong and Wang, Jie},
  journal={Nano Letters},
  volume={26},
  number={34},
  pages={11514--11520},
  year={2026},
  publisher={ACS Publications}
}

@article{PhysRevB.74.144401,
  title = {Distribution of antiferromagnetic spin and twin domains in NiO},
  author = {S{\"a}nger, I. and Pavlov, V. V. and Bayer, M. and Fiebig, M.},
  journal = {Phys. Rev. B},
  volume = {74},
  issue = {14},
  pages = {144401},
  numpages = {9},
  year = {2006},
  month = {Oct},
  publisher = {American Physical Society},
  doi = {10.1103/PhysRevB.74.144401},
  url = {https://link.aps.org/doi/10.1103/PhysRevB.74.144401}
}

@article{PhysRevLett.105.077402,
  title = {Spin Oscillations in Antiferromagnetic NiO Triggered by Circularly Polarized Light},
  author = {Satoh, Takuya and Cho, Sung-Jin and Iida, Ryugo and Shimura, Tsutomu and Kuroda, Kazuo and Ueda, Hiroaki and Ueda, Yutaka and Ivanov, B. A. and Nori, Franco and Fiebig, Manfred},
  journal = {Phys. Rev. Lett.},
  volume = {105},
  issue = {7},
  pages = {077402},
  numpages = {4},
  year = {2010},
  month = {Aug},
  publisher = {American Physical Society},
  doi = {10.1103/PhysRevLett.105.077402},
  url = {https://link.aps.org/doi/10.1103/PhysRevLett.105.077402}
}

@article{das_evolution_2018,
  title = {On the evolution of antiferromagnetic nanodomains in {NiO} thin films: {A} {LEEM} study},
  volume = {449},
  issn = {0304-8853},
  url = {https://www.sciencedirect.com/science/article/pii/S0304885317315809},
  doi = {https://doi.org/10.1016/j.jmmm.2017.10.060},
  journal = {Journal of Magnetism and Magnetic Materials},
  author = {Das, Jayanta and Menon, Krishnakumar S. R.},
  year = {2018},
  pages = {415--422},
}

@article{polewczyk2025control,
  title={Control of the antiferromagnetic domain configuration and N{\'e}el axis orientation with epitaxial strain},
  author={Polewczyk, Vincent and Petrov, Aleksandr Yu and Sarpi, Brice and Backes, Dirk and Elnaggar, Hebatalla and Wadhwa, Payal and Filippetti, Alessio and Rossi, Giorgio and Torelli, Piero and Vinai, Giovanni and others},
  journal={Communications Materials},
  volume={6},
  number={1},
  pages={153},
  year={2025},
  publisher={Nature Publishing Group UK London}
}

@article{zhu2024structural,
  title={Structural degeneracy and formation of crystallographic domains in epitaxial LaFeO3 films revealed by machine-learning assisted 4D-STEM},
  author={Zhu, Menglin and Lanier, Joseph and Flores, Jose and da Cruz Pinha Barbosa, Victor and Russell, Daniel and Haight, Becky and Woodward, Patrick M and Yang, Fengyuan and Hwang, Jinwoo},
  journal={Scientific Reports},
  volume={14},
  number={1},
  pages={4198},
  year={2024},
  publisher={Nature Publishing Group UK London}
}

\section*{End matter}
\appendix
\renewcommand{\thefigure}{A\arabic{figure}} %
\renewcommand{\thetable}{A\Roman{table}}

% Change numbering format
\setcounter{figure}{0} % Reset figure counter
\setcounter{table}{0} % Reset table counter

%\ad{Probably should just move everything to the SI (or everything to the appendix)}
The anharmonic energy surface was parametrized by fitting a $11\times11\times11$ grid of
frozen-phonon DFT total energies to the biquadratic model
\begin{align}
  E = \; & E_0
    + \tfrac{1}{2}\Omega_\mathrm{IR}(Q_\mathrm{IR,a}^2 + Q_\mathrm{IR,b}^2)
    + \tfrac{1}{2}\Omega_\mathrm{R} Q_\mathrm{R}^2 \nonumber \\
         & + \tfrac{1}{2}\lambda Q_\mathrm{IR,a}^2 Q_\mathrm{IR,b}^2
    + \tfrac{1}{2}g(Q_\mathrm{IR,a}^2 + Q_\mathrm{IR,b}^2)Q_\mathrm{R}^2 \nonumber \\
         & + \gamma Q_\mathrm{IR,a} Q_\mathrm{IR,b} Q_\mathrm{R}^2\,,
  \label{eq:energy_surface}
\end{align}
where $\Omega_\mathrm{IR}$ and $\Omega_\mathrm{R}$ are the harmonic mode stiffnesses, $\lambda$ is the
IR-IR biquadratic coupling, $g$ is the biquadratic IR-Raman coupling coefficient, and $\gamma$
is a secondary mixed term. The fit achieves $R^2 = 0.999$ and RMSE $= 3.8\times10^{-5}$\,eV;
the resulting parameters are listed in \autoref{tab:fit_coeffs} and additional details are provided in the SM \cite{supp}.

\begin{table}[h]
  \centering
  \caption{Parameters of the biquadratic energy-surface fit, \autoref{eq:energy_surface}, for NiO $\Gamma_3^-(10,11)$ modes in the standard mass-weighted normal coordinate.}
  \label{tab:fit_coeffs}
  \begin{ruledtabular}
  \begin{tabular*}{\columnwidth}{lll}
  Parameter & Value & Unit \\
\hline
  %$E_0$          & $-20.751$            & eV \\
  $\Omega_\mathrm{IR,a}$     & $0.506$             & eV\,\rm \AA$^{-2}$\,\rm amu$^{-1}$ \\
  $\omega_\mathrm{IR,a}$&11.13& THz\\
  $\Omega_\mathrm{IR,b}$     & $0.506$             & eV\,\rm \AA$^{-2}$\,\rm amu$^{-1}$ \\
  $\omega_\mathrm{IR,b}$&11.13& THz\\
  $\Omega_\mathrm{R}$     & $0.975$             & eV\,\rm \AA$^{-2}$\,\textrm{amu}$^{-1}$ \\
  $\omega_\mathrm{R}$&15.44& THz\\
  $\lambda$ & $1.33\times10^{-2}$   & eV\,\AA$^{-4}$\,\textrm{amu}$^{-2}$ \\
  %$\lambda_{13}$ & $6.65\times10^{-3}$   & eV\,\AA$^{-4}$\,amu$^{-2}$ \\
  %$\lambda_{23}$ & $6.65\times10^{-3}$   & eV\,\AA$^{-4}$\,amu$^{-2}$ \\
  $g$ & $1.06\times10^{-1}$   & eV\,\AA$^{-4}$\,amu$^{-2}$ \\
  $\gamma$& $1.00\times10^{-4}$   & eV\,\AA$^{-4}$\,amu$^{-2}$ \\
  \end{tabular*}
  \end{ruledtabular}
\end{table}

\end{document}